\documentclass[aps,pra,twocolumn,a4paper,groupedaddress,floatfix,10pt,longbibliography]{revtex4-1}
\usepackage[utf8]{inputenc}
\usepackage{amsmath}
\usepackage{amsfonts}
\usepackage{amssymb}
\usepackage{graphicx}
\usepackage[T1]{fontenc}

\def\ket#1{\left|#1 \right\rangle}

\begin{document}

\title{Bosonic Fractional Quantum Hall States on a Finite Cylinder}
\author{Paolo Rosson${}^{1}$}
\author{Michael Lubasch${}^{1}$}
\author{Martin Kiffner${}^{2,1}$}
\author{Dieter Jaksch${}^{1,2}$}

\affiliation{Clarendon Laboratory, University of Oxford, Parks Road, Oxford OX1
3PU, United Kingdom${}^1$}
\affiliation{Centre for Quantum Technologies, National University of Singapore,
3 Science Drive 2, Singapore 117543${}^2$}
\date{\today}

\begin{abstract}
We investigate the ground state properties of a bosonic  Harper-Hofstadter model with local interactions on a finite cylindrical lattice with filling fraction $\nu=1/2$. We find that our system supports topologically ordered states by calculating the topological entanglement entropy, and its value is in good agreement with the theoretical value for the $\nu=1/2$ Laughlin state. By exploring the behaviour of the density profiles, edge currents and single-particle correlation functions, we find that the ground state on the cylinder shows all signatures of a fractional quantum Hall state even for large values of the magnetic flux density. 
Furthermore, we determine the 
dependence of the correlation functions 
and edge currents on the interaction strength.  We find 
that depending on the magnetic flux density, the transition towards Laughlin-like behaviour can 
be either smooth or happens abruptly for some critical 
interaction strength.
\end{abstract}

\maketitle
\section{Introduction} \label{Intro}

An exciting approach for gaining novel insights into the physics of strongly
correlated many-body quantum systems is provided by  the concept of quantum simulation~\cite{cirac:12,Johnson:14}. 
A particularly promising platform for the simulation of many-body quantum
lattice systems are ultracold atoms trapped in periodic electromagnetic fields~\cite{jaksch:98,greiner:02,bloch:08,Joerdens2008,gullans:12}. 
In state-of-the-art optical lattice experiments
one can control, manipulate and observe atoms with single-site resolution~\cite{bakr:09,bakr:10,sherson:10,weitenberg:11,gericke:08,wuertz:09, Cheuk2015}
which opens up unprecedented possibilities for the realization and observation of 
novel physical phenomena. 
Quantum Hall systems are a fundamental paradigm in condensed matter physics~\cite{Klitzing1980,FQH,Tsui1982}, and thus the experimental realization  of these systems 
in optical lattices  has attracted tremendous interest in recent years~\cite{artgauge,ExpHH, Miyake2013,ExpChern}. Of particular interest are 
fractional quantum Hall states~\cite{cooper:13,FQH,Cooper2008, Harper2014, Scaffidi2014} which require interactions 
between the particles~\cite{Tai2017}  and are thus difficult to realize. 

Fractional quantum Hall systems on lattices are described by an extension of the Harper-Hofstadter (HH) model~\cite{Hofstadter} in order to account for interactions~\cite{Harper1955,Hofstadter,Hafezi2007}. 
The realization of fractional quantum Hall states requires appropriate filling fractions 
$\nu=n_b/n_{\phi}$~\cite{FQH,Laughlin1983}, where 
$n_{\phi}$ is the density of magnetic fluxes per plaquette and $n_b$ is the 
density of particles. 
In small lattice systems permitting exact diagonalisation (ED) calculations~\cite{Sorensen2005, Hafezi2007, Moeller2009, Sterdyniak2012} the properties of the ground state can be simply analyzed by calculating the overlap with the corresponding Laughlin wave function. 

Larger-scale systems become theoretically accessible with density matrix renormalization group (DMRG) techniques~\cite{DMRG, White1993}, but require alternative methods for characterising the ground state features. 
For example,  the presence of a gapped, insulating bulk 
and gapless edge modes in a fractional quantum Hall state 
gives rise to a uniform density in the bulk and a density spike 
near the edges~\cite{CIFTJA2011, Rezayi1994}, chiral edge currents~\cite{Halperin1982, XIAO-GANG1992} and  algebraically decaying one-particle correlation functions near the edge~\cite{He2017, Gerster2017}. 
Furthermore, an unambiguous indication of topological order of the ground state 
is the topological entanglement entropy, which can be calculated
for Laughlin states~\cite{TEE,Fendley2007,Dong2008}. 
These observables have been used to characterize fractional quantum Hall states  on a torus, a square lattice~\cite{Gerster2017, He2017,Raciunas2018} and ladder 
geometries~\cite{Budich2017, Strinati2017, Petrescu2017}. 
Other microscopic models supporting Laughlin states have been considered \cite{Cincio2013, Nielsen2013, Bauer2014} and a number of methods to characterize topologically ordered ground states have been proposed such as modular matrices and momentum polarization~\cite{Cincio2013,tu2013,zaletel2013}.

Recently, a theoretical proposal for the realization of cylindrical lattices has been put forward~\cite{Lacki2016}. This geometry is particularly well suited for investigating manifestations of gapless edge modes due  to the two system boundaries in one direction and periodic boundary conditions (PBC) in the other direction. Furthermore,  it has been shown  that the cylinder geometry can support fractional quantum Hall states for Fermions and small system sizes using ED techniques~\cite{Lacki2016}.

Here we investigate ground states of  the interacting HH model for bosons on a cylinder geometry with DMRG methods.
We focus on states with filling 
fraction $\nu=1/2$ and show that the cylinder geometry supports topologically 
ordered states. 
More specifically, we calculate the topological entanglement entropy  for a system with magnetic flux density $n_{\phi}=1/6$ and find that its 
value  is  very close to that of a $\nu=1/2$ Laughlin state in the continuum. 

Our DMRG approach  allows us to study  how  the density profile, the edge currents and single-particle correlation functions scale with the length of the cylinder and the magnitude of the on-site interaction. 
These calculations  are performed in  the as yet little explored regime of large flux densities where lattice effects become important.
We find that all observables of interest exhibit the characteristic features of the  Laughlin state  in the case of large system sizes and  hard-core bosons. 

Finally, we address the dependence of the correlation functions and edge currents on the on-site interaction strength $U$  and find that the Laughlin-like behaviour is adapted continuously with increasing $U$ for $n_{\phi}=1/3$. 
On the other hand, we show that a slightly smaller flux density  $n_{\phi}=1/4$ can  give rise to an abrupt increase in the importance of the edge currents relative to the bulk.

This paper is organized as follows.  In Sec.~\ref{Phys_sys} 
we describe the geometry of the system and introduce the theoretical model. 
Our results are presented in Sec.~\ref{Results}. 
The topological entanglement entropy is discussed  in Sec.~\ref{TEE}, 
and the density profiles,  currents and  correlation functions are covered in Sec. \ref{Observb}. The dependence of the correlation functions and edge currents 
on  the interaction strength is presented in  
Sec.~\ref{BH_case}. Finally, a discussion and summary of  
our results is provided in Sec.~\ref{Conclusions}.
\begin{figure}[!t] 
\begin{center}
\includegraphics[width=1\linewidth]{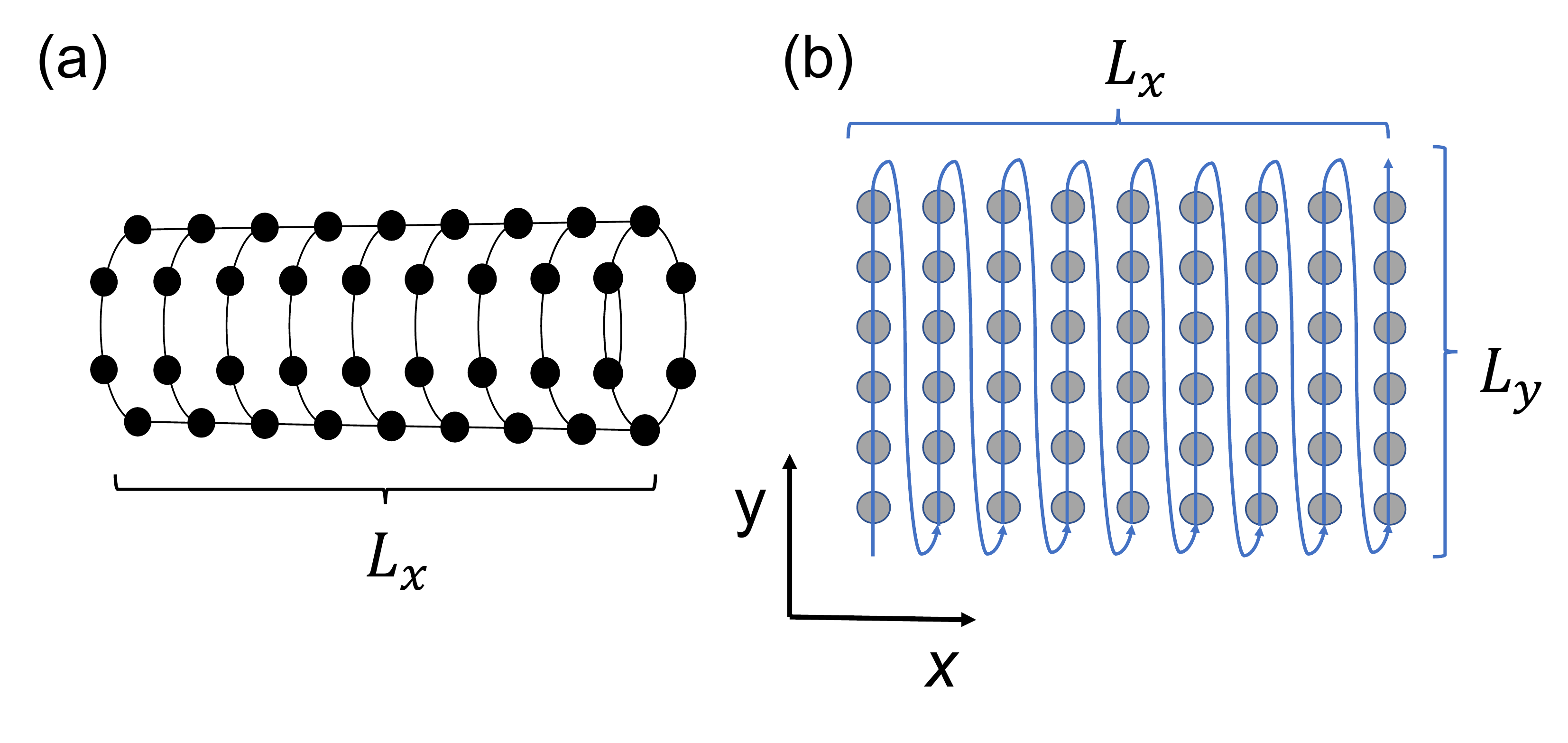}
\end{center}
\caption{(Color online)    (a) The system of interest is a cylindrical lattice 
with length $L_x$ and  circumference $L_y$. 
(b) Rectangular lattice corresponding to the cylindrical lattice in (a). 
Threading through  the  two-dimensional lattice with $L_x\times L_y$ sites as indicated by the blue line allows one to 
map the system to a one-dimensional chain with long-range hopping terms. }
  \label{lattice}
\end{figure}
\section{Model} \label{Phys_sys}
 We consider the interacting HH model for spinless bosons on a square lattice with $L_x$ sites in the $x$-direction and $L_y$ sites in the $y$-direction. We impose PBC along the $y$ direction, turning the lattice into a   cylinder as shown in Fig.~\ref{lattice}. The bosons are subject to a gauge field giving rise to a uniform magnetic field orthogonal to the lattice. The Hamiltonian of the system is
\begin{align}
 H = H_{\text{kin}} + H_{U}\,, \label{Ham}
\end{align}
where $H_{\text{kin}}$ and $H_{U}$ describe the single-particle kinetic energy and the on-site interaction, respectively. We find 
\begin{align}
H_{\text{kin}} = & 
-J \left[\sum_{x=1}^{L_x-1}\sum_{y=1}^{L_y} c^{\dagger}_{x,y}c_{x+1,y} \right.\notag \\
 & \hspace*{1cm}
 + \sum_{x=1}^{L_x} \sum_{y=1}^{L_y-1} e^{i2\pi x\phi} c^{\dagger}_{x,y}c_{x,y+1}\notag \\
& 
\hspace*{1cm} 
\left. + \sum_{x=1}^{L_x}  e^{i2\pi x\phi}c^{\dagger}_{x,L_y}c_{x,1} \right] 
+ \text{H.c.} \,,\label{Hkin}
\end{align}
where $c_{x,y}$ is the annihilation operator for a boson at a site labelled by the coordinates $(x,y)$  and satisfying bosonic commutation relations $[c_i, c_j^\dagger]=\delta_{ij}$ and   $J$ is the hopping amplitude. 
Note that $H_{\text{kin}}$ in Eq.~(\ref{Hkin}) is written in the Landau gauge. 
When a particle hops around a plaquette it picks up a phase $e^{i2\pi \phi}$, and hence the number of flux quanta per  plaquette is  $n_{\phi}\equiv\phi$.
The interaction term $H_U$ in Eq.~(\ref{Ham}) is defined as 
\begin{align}
H_U =  U \sum_{x=1}^{L_x}\sum_{y=1}^{L_y}n_{x,y} (n_{x,y}-1)\,, \label{Hint}
\end{align}
where $U$ is  the on-site interaction strength and $n_{x,y}=c^{\dagger}_{x,y}c_{x,y}$ is the number operator at site $(x,y)$.  
We truncate the maximal occupation number of each site at $n_{\text{max}}$ in our numerical calculations. The required value of $n_{\text{max}}$ in order to achieve convergence depends on the interaction strength as specified in Sec.~\ref{Results}. 
We also consider the hard-core boson limit $U/J\rightarrow\infty$ which can be 
realised by setting $n_{\text{max}} =1$.  

We obtain the ground state of the system by performing DMRG calculations.   To this end, we  map  the 2D lattice with PBC
to a 1D lattice~\citep{2D} as shown in Fig.~\ref{lattice}(b). While the 2D model in Eq.~(\ref{Ham}) comprises only nearest-neighbour (NN) hopping terms, the mapping to 
1D  introduces  hopping terms with range up to $L_y$.  
In order to efficiently treat these long-range hopping terms, we  build the matrix product operator (MPO) that describes the 2D Hamiltonian with cylindrical boundary conditions using the finite automata technique~\cite{automata}. In this method, the MPOs are derived from a complex weighted finite automaton as described 
in Appendix~\ref{appendixA}. All our DMRG calculations are carried out with the TNT library~\cite{TNT} and use $U(1)$ number conservation symmetry. 
\section{Results} \label{Results}
Here we present our DMRG results for the ground state on a cylinder with filling 
fraction $\nu=1/2$ and different values of $\phi$. 
We begin with a discussion of  the topological entanglement entropy for 
a system with $\phi = 1/6$  in Sec.~\ref{TEE}. Our calculations 
for  the particle density, currents and  correlation functions are presented in 
Sec.~\ref{Observb} for $\phi = 1/3$.  
All calculations in Secs.~\ref{TEE} and~\ref{Observb} correspond to the 
hard-core boson limit. This condition is relaxed in 
Sec.~\ref{BH_case}, where we consider variable interaction strengths $U/J$. 
In particular, we present calculations of the bulk and edge currents 
for variable interaction strengths and two different magnetic flux densities. 
\subsection{Topological Entanglement Entropy} \label{TEE}
\begin{figure}[t!]
  \begin{center}
    \includegraphics[width=\linewidth]{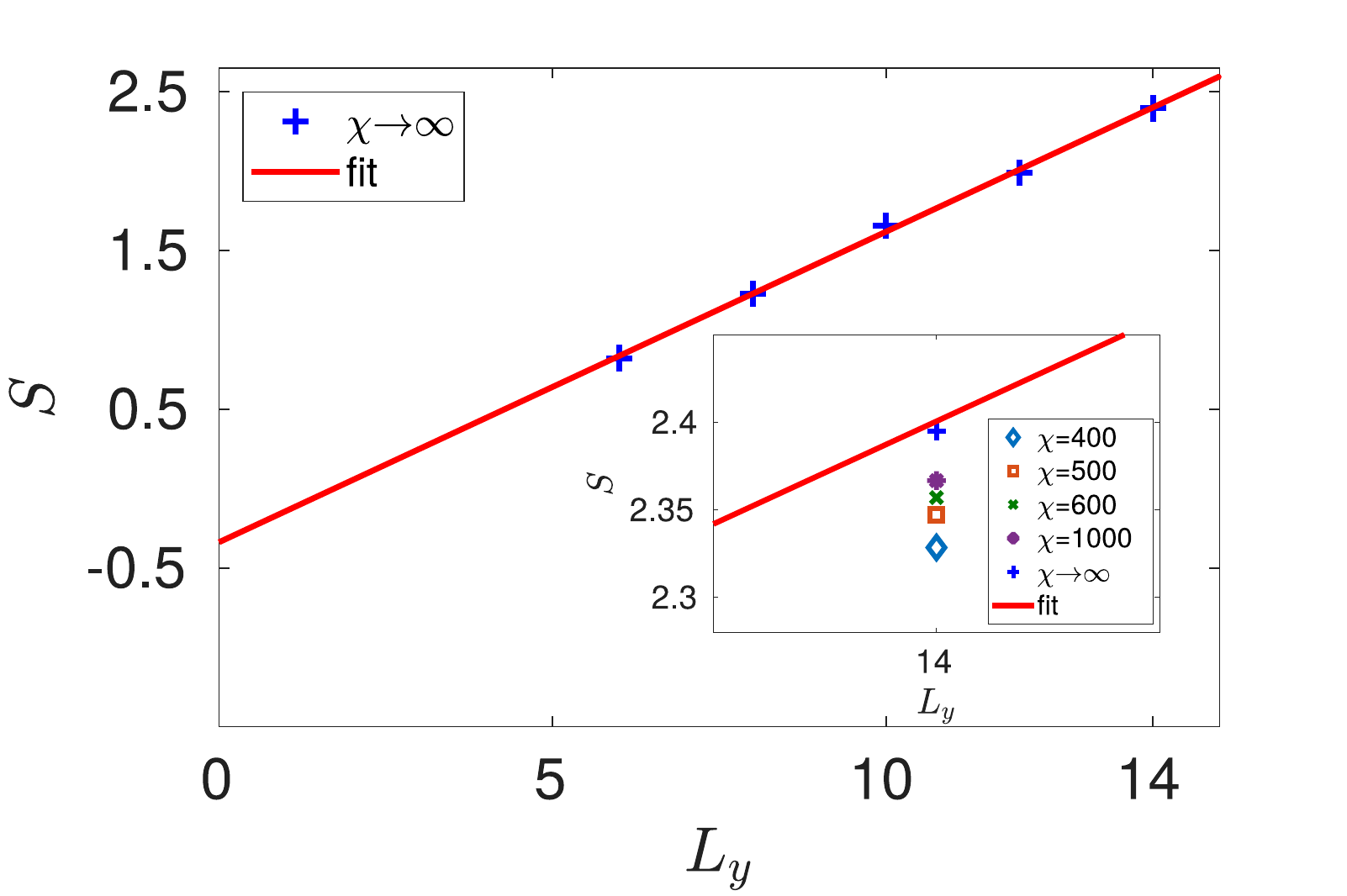}
  \end{center}
	\caption{  (Color online)  Entanglement entropy $S$ for a cylinder of 
	length $L_x = 6$,  filling fraction $\nu=1/2$ and $\phi=1/6$ as a function of 
	 circumference $L_y$.  
	  The blue crosses are the results of 
	 an extrapolation of $S$ to $\chi\rightarrow \infty$, and the red solid line is the linear fit to these values. The inset shows the data points obtained for different bond dimensions $\chi$ at $L_y = 14$. } 
  \label{fig:TEE}
\end{figure}
The topological order of quantum many-body states 
can be quantified using the topological entanglement entropy  $\gamma$ 
as shown in~\cite{TEE}. A non-zero value of $\gamma>0$  unambiguously   
indicates  topological order of the ground state and the presence 
of anyonic quasi-particle excitations. 
The topological entanglement entropy can be calculated from a correction to the area 
law~\cite{Vidal2003,Eisert2010} for the entanglement entropy  
\begin{align}\label{eq:TEE}
S(L)= cL-\gamma\,,
\end{align}
where  $L$ is the length of the system boundary created by dividing the system into two  parts, and  $c$ is a non-universal constant. 
DMRG calculations lend themselves particularly well for computing $\gamma$ 
since $S$ can be evaluated from the Schmidt values $\lambda_i$ 
entering the matrix product representation of the ground state, 
\begin{align}
S=-\text{tr}[\rho_L\ln(\rho_L)]=-\sum_i |\lambda_i|^2 \ln(|\lambda_i|^2)\,,
\end{align}
where $\rho_L$ is the density matrix of the left part of the system after a cut. 
We compute $S$ for the ground state  in the cylinder
geometry with filling fraction $\nu = 1/2$ and $\phi = 1/6$ for a cylinder of length $L_x=6$ and different values of the circumference $L_y$ as shown in Fig.~\ref{fig:TEE}.
We perform the cut by dividing the cylinder in two equal parts, each consisting of  $3\times L_y$ sites. 
Our DMRG calculations lead to a two-fold degenerate ground state for the considered system sizes due to the centre-of-mass degeneracies in the corresponding system on a torus~\cite{Kol1993}. In using Eq.~(\ref{eq:TEE}) for calculating the topological entanglement entropy we thus assume that DMRG automatically selects the minimally entangled states~\cite{jiang2012}, and that only Abelian anyons are present in the system~\cite{zhang2012}.

We calculate $\gamma$ by extrapolating all data points for 
$S$ to $\chi\rightarrow\infty$ resulting in $S_{\infty}(L_y)$. We then perform a linear fit to the values $S_{\infty}(L_y)$ and obtain
\begin{align}
 \gamma\approx 0.34\pm 0.05 \,,
\end{align}
 where the error represents the 68\%  confidence interval of the linear fit 
 to the $S_{\infty}$ values. This clearly demonstrates 
 that the cylinder geometry supports topologically ordered ground states. Moreover, 
 the value of $\gamma$ is in  good agreement with the expected value of 
 $\gamma =\ln\sqrt{2}\approx 0.347$ 
 for a Laughlin state with filling fraction $\nu=1/2$~\cite{Fendley2007,Dong2008}. 
\subsection{Particle density, edge currents and correlation functions} \label{Observb}
 Here we present results for  the particle density, the edge currents and the correlation functions of our system. More specifically, we consider the ground state of our cylindrical lattice  in the hard-core boson limit and focus on the state with filling fraction $\nu=1/2$ and $\phi=1/3$.
\begin{figure}[t] 
  \begin{center}
    \includegraphics[width=1\linewidth]{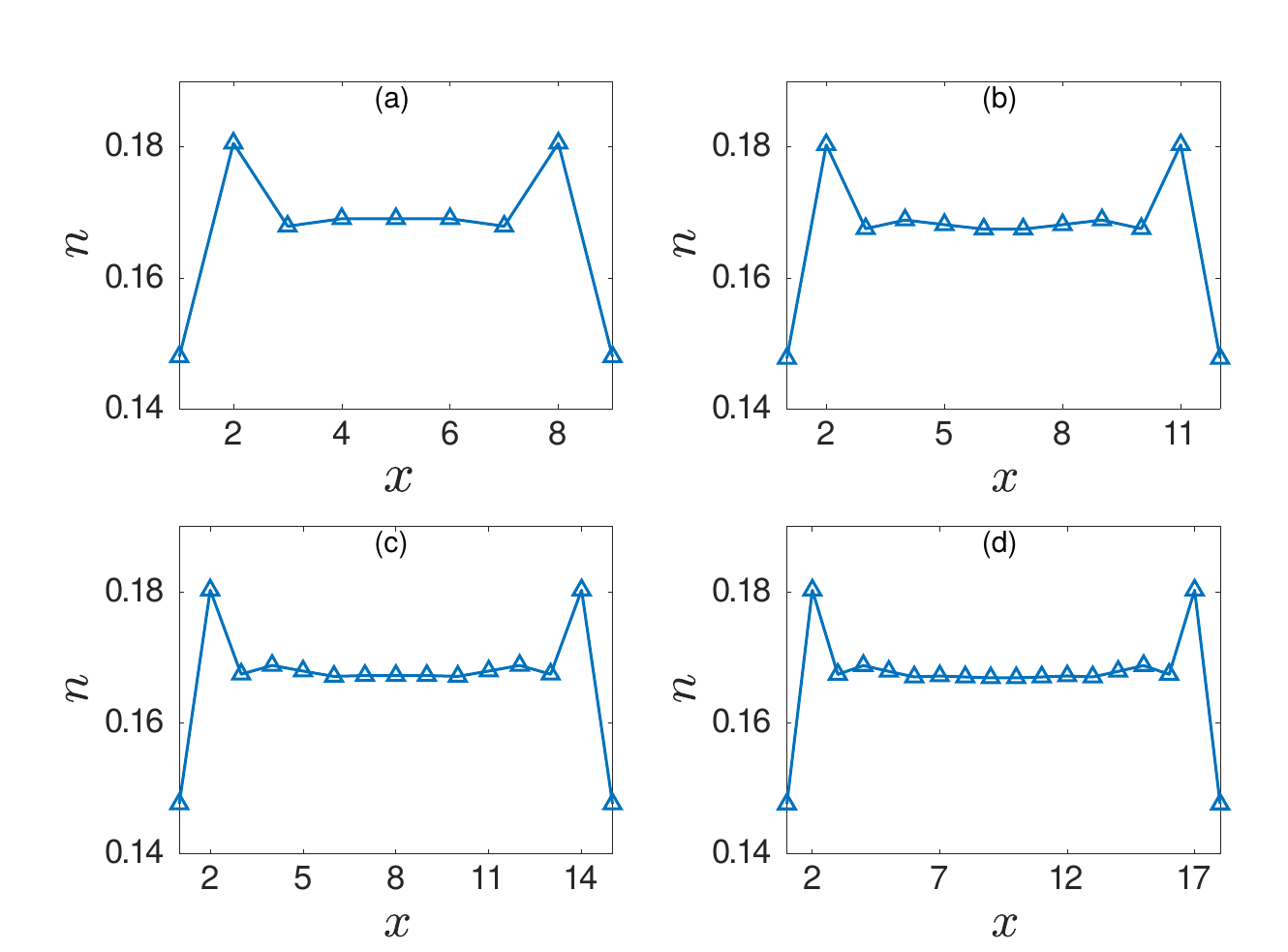} 
  \end{center}
  \caption{\label{density_nu12} (Color online) Particle density $n$  as a function of the $x$ coordinate for system sizes $L_x\times L_y$: (a) $9\times6$,  (b) $12\times6$, (c) $15\times6$  and (d) $18\times6$. $n$ is dimensionless. Lines connecting the data points are a guide to the eye, and all calculations have been performed with bond dimension $\chi=1000$. }
\end{figure} 
Throughout this section, we consider a cylinder with circumference $L_y=6$ and four
different values of $L_x$. These values are chosen such that the ground state 
is non-degenerate in order to obtain unambiguous results for the observables of interest. System sizes leading to non-degenerate ground states can be found by first considering  a torus where  unique ground states can be obtained for specific values of $L_x$ and $L_y$~\cite{Kol1993}. 
For all cases investigated here the corresponding  cylinder geometry also features a non-degenerate ground state. 
\begin{figure}[t!] 
  \begin{center}
    \includegraphics[width=1\linewidth]{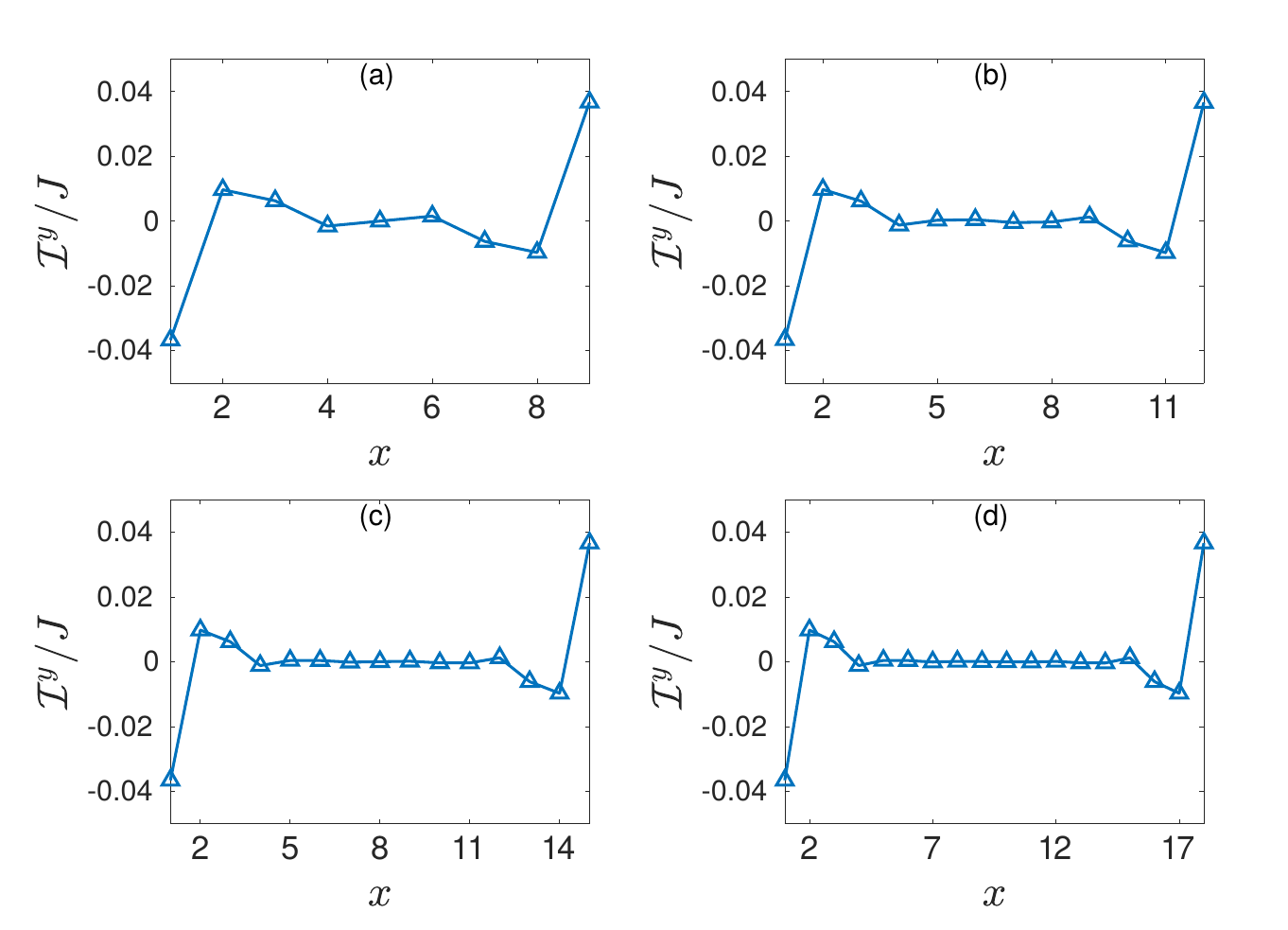} 
  \end{center}
  \caption{\label{current_nu12}  (Color online)  Current around the cylinder $\mathcal{I}^{y}$ as a function of the $x$ coordinate for system sizes $L_x\times L_y$: (a) $9\times6$,  (b) $12\times6$, (c) $15\times6$  and (d) $18\times6$. 
  Lines connecting the data points are a guide to the eye, and all calculations are for bond dimension $\chi=1000$. }
\end{figure} 

We start by examining the local density defined as 
\begin{align}
n(x)=\frac{1}{L_y}\sum\limits_{y=1}^{L_y}\langle c^\dagger_{x,y}c_{x,y}\rangle\,,
\label{ndef}
\end{align}
where we take the average over the circumference because of the  
rotational symmetry of the cylinder. 
The density profile of the ground state 
is shown in Fig.~\ref{density_nu12} for four different lengths of the cylinder. In all cases, the density of the central ring is $n\approx0.17\approx n_b$ ($n_b$: mean number of bosons), and the density profile exhibits a pronounced spike at the second ring.  As the length of the cylinder increases, the density profile increasingly resembles the density profile of the corresponding Laughlin state in the continuum~\cite{CIFTJA2011, Rezayi1994}. 

Next, we investigate the mean currents $\mathcal{I}^{y}$ flowing around the cylinder in $y$-direction, 
\begin{align}
\mathcal{I}^{y}(x)= \frac{1}{L_y}\sum\limits_{y=1}^{L_y}\langle i J e^{i2\pi x\phi}c_{x,y}^{\dagger} c_{x,y+1} + \text{H.c.} \rangle\,,
 \label{currdef}
\end{align}
where the average is taken over the circumference of the cylinder and site $L_y + 1$ corresponds to site $1$  due to the rotational symmetry of the cylinder. 
 The currents for different $x$ coordinates and variable cylinder lengths are shown in Fig.~\ref{current_nu12}. We find that the current is dominant at the edges and takes on its smallest values near the center of the cylinder where the currents fluctuate around zero. As the length of the cylinder increases, the current fluctuations in the bulk become smaller. Note that the currents flow in opposite directions at the two edges of the cylinder, which signifies their  chiral nature. Our results for the currents are consistent with gapless edge states and an insulating bulk and  are qualitatively similar  to those for a $\nu=1/2$ Laughlin state. 

\begin{figure}[t!] 
    \begin{center}
    \includegraphics[width=0.8\linewidth]{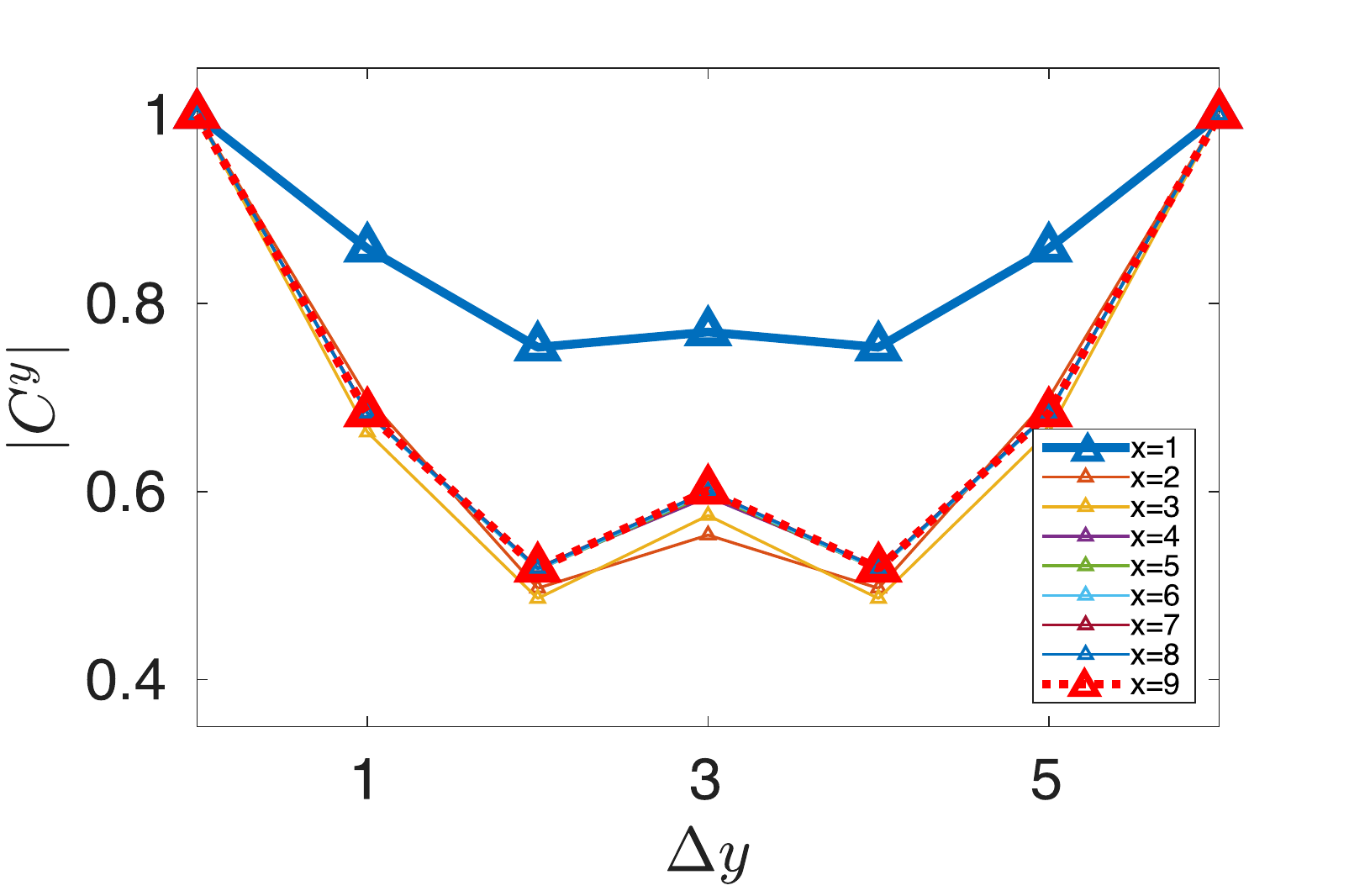} 
    \end{center}
\caption{\label{fig:correl_nu12} (Color online) (a) Correlation functions $C^y$ for system size $L_x\times L_y$: $18\times6$. Each line represents the correlation function around a ring identified by its $x$-coordinate. Only the rings from $x=1$ to $x=L_x/2$ are shown 
due to the inversion symmetry of the cylinder. 
 Lines connecting the data points represent a guide to the eye and all 
 calculations are for bond dimension $\chi=1000$. 
}
\end{figure}

Finally we consider the (normalized) one-particle correlation functions along the $y$-direction, 
\begin{align}
 C^y(x,\Delta y)=\frac{\langle c_{x,1}^{\dagger} c_{x,1 + \Delta y}\rangle}{ n(x) }.
 \label{Cdef}
\end{align}
The correlation function $C^y(x,\Delta y)$ on ring $x$ is a measure for correlations 
between two sites that are $\Delta y$ apart in the radial direction. 
The correlation functions  for a cylinder of length $L_x=18$ and circumference $L_y=6$
are shown in Fig.~\ref{fig:correl_nu12}. We find that the correlation function on the first ring ($x=1$) stands out from all others since it decays  
significantly slower  with $\Delta y$ than all other correlation functions. 
Note that a qualitatively similar behaviour of the correlation functions is found for all other system sizes investigated in Figs.~\ref{density_nu12} 
and~\ref{current_nu12}. 
Our results are consistent with an algebraic decay of correlation functions at the edge and an exponential decay in the bulk, as expected for a fractional quantum Hall state with gapless edge modes and gapped bulk excitations. 
\subsection{Interaction-strength dependence} \label{BH_case}
All results in Sec.~\ref{Observb}  were obtained for hard-core bosons, i.e., infinitely strong on-site interactions. Since  interactions in the condensed matter systems of interest are not infinitely strong, the question arises how finite interaction strengths modify these results. Here we address this question 
and calculate the  correlation functions and currents for a cylindrical lattice 
of length $L_x = 9$ and circumference $L_y=6$, $\phi=1/3$ and $\nu=1/2$ 
for variable interaction strengths $U/J$. 

\begin{figure}[t!] 
    \begin{center}
    \includegraphics[width=1\linewidth]
    {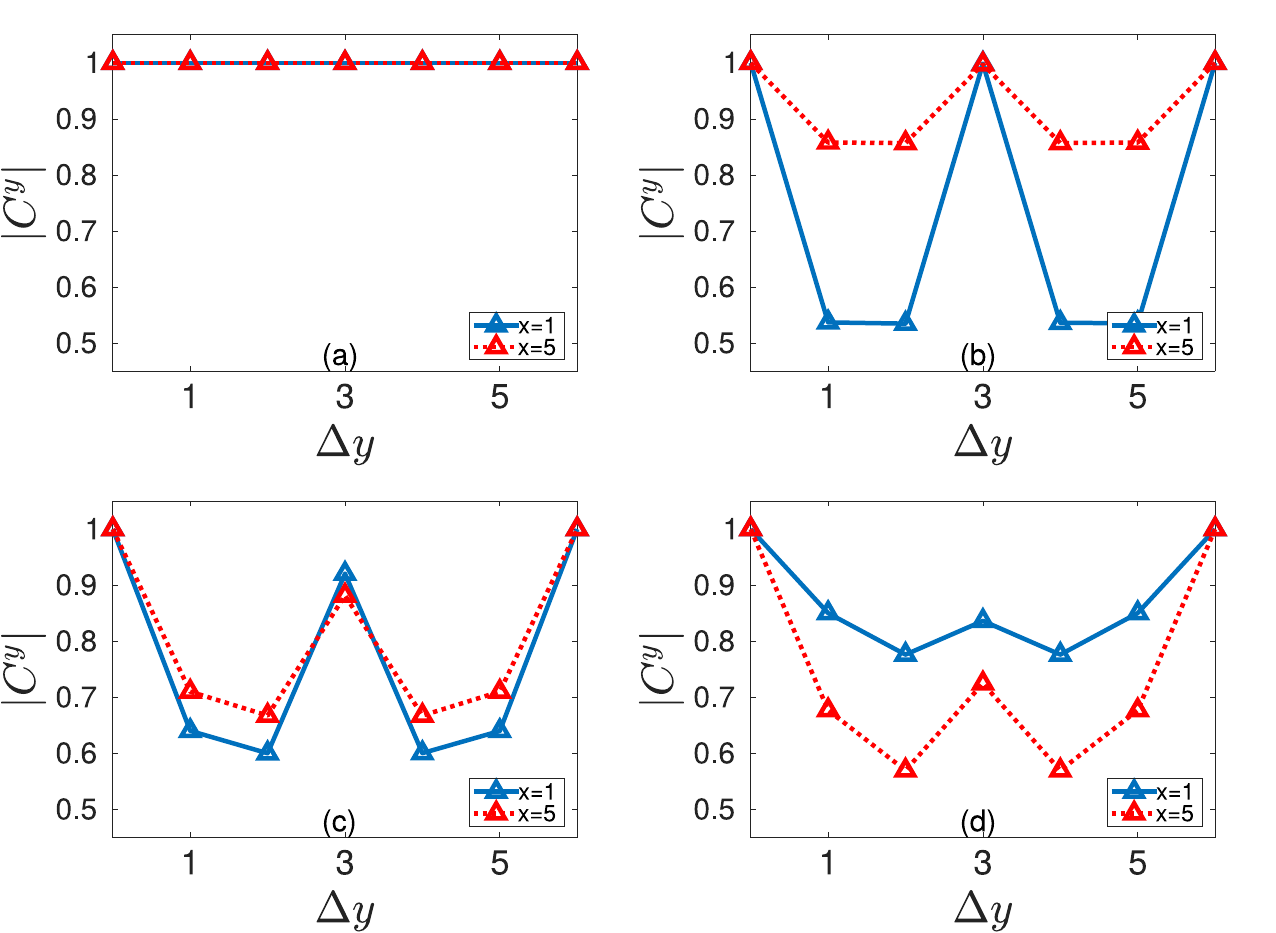} 
  \end{center}
  \caption{\label{BHplots} (Color online) Correlation functions $C^y$   for an $L_x \times L_y=9\times6$ lattice, with $\nu=1/2$  and $\phi=1/3$. 
  The interaction strength takes on values (a) $U/J=0$, (b) $U/J=0.1$, (c) $U/J=2$, (d) $U/J=8$.  Lines connecting the data points are a guide to the eye. 
  The maximal occupation number at each site is $n_{\text{max}}=4$ and the bond dimension is $\chi=500$.  For $U=0$ we used $n_{\text{max}}=9$, which is equal to the total number of particles in the system.}
\end{figure}

\begin{figure}[t!] 
  \begin{center}
     \includegraphics[width=1\linewidth]{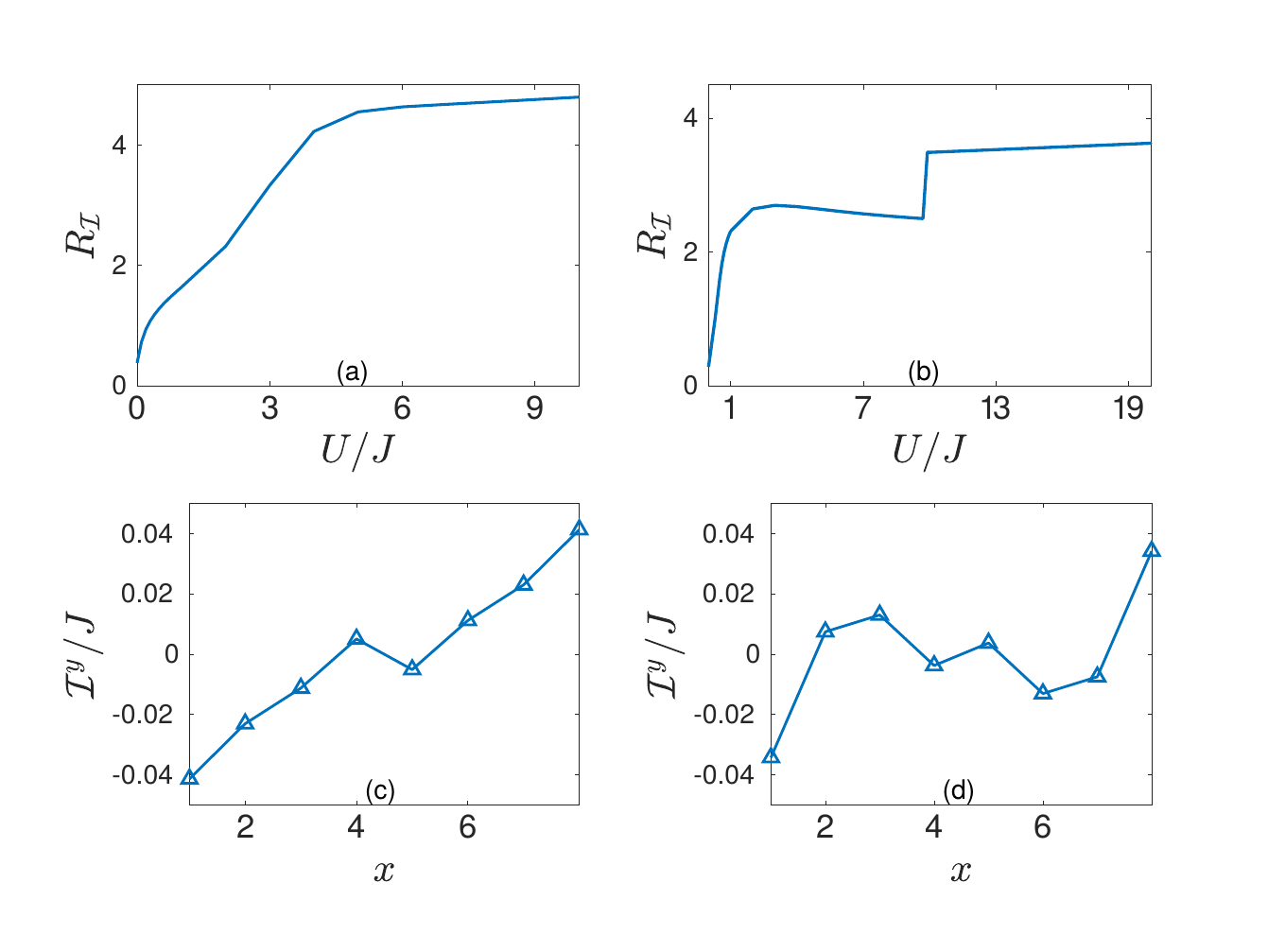}
  \end{center}
  \caption{\label{BHcurr} (Color online) 
(a)   $R_{\mathcal{I}}$ as defined in Eq.~(\ref{Crat}) for a system of size  
$L_x\times L_y=9\times 6$, $\phi=1/3$ and $\nu=1/2$. 
 (b) $R_{\mathcal{I}}$ for a system of size  
$L_x\times L_y=8\times 4$, $\phi=1/4$ and $\nu=1/2$. 
 For the DMRG calculations, we  used a maximum occupation number per site $n_{\text{max}}=4$, and bond dimension up to $\chi=300$.
 The current $\mathcal{I}^{y}(x)$ for the system in (b) is shown in 
 (c) and (d)  for $U/J = 9.6$  and $U/J = 9.9$, respectively. 
 Lines connecting the data points  in (c) and (d) are a guide to the eye. 
  }
\end{figure}

We begin with a discussion of  the correlation functions $ C^y(x,\Delta y)$ defined in Eq.~(\ref{Cdef}). In the non-interacting case $U=0$ we can find an  analytical solution of the ground state which is presented in Appendix~\ref{appendixB}. This exact solution implies 
that all correlation functions are constant, see  Fig.~\ref{BHplots}(a). 
Our numerical results for   $C^y(x,\Delta y)$ with $x=1$ and for the central ring ($x=5$) are  shown in Figs.~\ref{BHplots}(b-d) for $U>0$. 
For small values of $U/J$, the correlations 
at the edge decay faster than those in the bulk [see Figs.~\ref{BHplots}(b) and~(c)]. 
Note that this is in stark contrast to the hard-core boson case where the correlation 
function at the edge decays slower than all other correlation functions, see Sec.~\ref{Observb}. 
We find that the decay of correlations at the edge becomes slower than that of the bulk correlations for $U/J\gtrsim 2.5$, and for $U/J=8$ the correlation functions are practically  indistinguishable from their hard-core boson counterparts as 
shown in Fig.~\ref{BHplots}(d).
These results are consistent with ED calculations for smaller systems, showing that for $\phi>0.2$ the interaction strength must exceed $U/J\gtrsim \phi $ in order 
to induce Laughlin-like states. 
In order to systematically investigate how the hard-core boson limit is approached with increasing interaction strength, we focus on the currents defined in Eq.~(\ref{currdef}) and introduce the quantity
\begin{align}
  R_{\mathcal{I}}= \frac{|\mathcal{I}^{y}(1)|}{\sigma(\text{bulk})}\,,
  \label{Crat}
\end{align}
where  $\mathcal{I}^{y}(1)$ is the edge current and $\sigma(\text{bulk})$ is  the standard deviation of the currents in the bulk.   The ratio $R_J$ quantifies the relative importance of the edge current  with respect to   the currents in the bulk. 
Our results for $R_J$ as a function of $U/J$ are shown in Fig.~\ref{BHcurr}. 
We find that the bulk currents dominate the edge currents for $U/J<0.5$, and the 
ratio $ R_{\mathcal{I}}$ increases monotonically with $U/J$ until 
it starts saturating  near $U/J\approx 5$. Similarly to the correlation functions, 
we thus find that Laughlin-like behaviour of the currents requires interaction strengths exceeding $U/J\gtrsim \phi$. 

A striking feature of Fig.~\ref{BHcurr}(a) is that the transition from zero to strong interactions is completely smooth and does not feature a sharp phase transition. 
Since a fractional quantum Hall state requires interactions, a sharp transition between  a state with no topological order for $U=0$ and a topologically ordered state for $U>0$ is expected in the continuum. This is confirmed by ED calculations for small systems~\cite{Hafezi2007} indicating that this topological 
phase transition happens at $U=0$ for small values of $\phi<0.2$. 
The reason for our observation of a smooth transition  is that we consider 
a large value of $\phi=1/3$ where there is a significant difference between the 
continuum and the lattice. The overlap between  the Laughlin state on the lattice and the actual ground state is small for  $U=0$ and increases gradually with increasing $U/J$~\cite{Hafezi2007}. This gradual increase of the overlap is consistent with our results for $R_{\mathcal{I}}$ in Fig.~\ref{BHcurr}(a).

A different situation should arise for smaller values of $\phi$ where the overlap with the Laughlin state at $U=0$ is larger~\cite{Hafezi2007}. In order to investigate this we consider a system with a smaller magnetic flux density $\phi=1/4$. Our results for $ R_{\mathcal{I}}$ are shown in Fig.~\ref{BHcurr}(b), showing that  $ R_{\mathcal{I}}$ increases sharply with $U/J$ until $U/J\approx 3$, where it levels off and subsequently decreases slightly. Most importantly, $ R_{\mathcal{I}}$ 
exhibits a  sharp jump at $U/J\approx9.8$. The corresponding currents  before and after the jump are shown in Figs.~\ref{BHcurr}(c) and~(d), respectively. 
For $U/J=9.6$ the  current grows approximately linearly with the position $x$ 
of the ring along the cylinder axis. 
On the contrary, for  $U/J=9.9$ the dependence of the current on  $x$ is 
similar to that of the hard-core boson case for $\phi = 1/3$, see 
Fig.~\ref{BHplots}(d). 
\section{Discussion and Summary \label{Conclusions}}
In this work, we have presented DMRG calculations for ground states 
of  an interacting bosonic Harper-Hofstadter model on a cylinder geometry. 
In order to apply DMRG methods to this system, we mapped the two-dimensional 
grid to a one-dimensional chain with long-range hopping terms and constructed 
the corresponding matrix product operators using the finite state automaton 
technique presented in~\cite{automata}.

We calculated the topological entanglement entropy for a filling fraction of $\nu=1/2$ 
and $\phi = 1/6$ and found that its value is larger than zero, see Sec.~\ref{TEE}. This shows that 
the cylinder geometry supports topologically ordered states even for relatively large values of the magnetic flux $\phi$. Moreover, the value of the topological entanglement entropy is very close to the theoretically expected result for 
a $\nu = 1/2$ Laughlin state.  

The cylinder geometry with PBC in one dimension and two edges is ideally suited for studying characteristic features of quantum Hall states which we present in Sec.~\ref{Observb}. 
We considered a filling fraction $\nu = 1/2$ and a relatively large magnetic flux density of $\phi=1/3$ that can be easily realised in cold atom experiments. 
It has previously been shown that the overlap between the Laughlin wave function on a grid and the actual ground state of small lattice systems is small for these parameters~\cite{Hafezi2007}. 
Nevertheless, we find that  the density profiles, currents and correlation functions for our states on the cylinder show all characteristic features of a fractional quantum Hall state described by a Laughlin function in the continuum. 
In particular, our results are consistent with a gapped bulk and gapless edge modes leading to chiral edge currents and algebraically decaying correlation functions.  

All our results in Secs.~\ref{TEE} and~\ref{Observb} were obtained in the hard-core 
boson limit. In order to address the dependence of our results on the on-site interaction strength, we evaluated the  correlation functions and currents as a function of the interaction strength in Sec.~\ref{BH_case}. We found that both observables differ 
significantly from those of a Laughlin state for small values of $U/J$. 
The currents and correlation functions start to show qualitative signatures 
of a Laughlin state for  $U/J>2.5$, and for $U\approx 8$ our results are very close to the hard-core boson limit. 
We systematically investigated the transition from weak to strong interactions for the ratio $R_{\mathcal{I}}$ between the edge current and  the current fluctuations  in the bulk. 
Our results do not show any signatures of a topological 
phase transition since $R_{\mathcal{I}}$ varies smoothly with $U/J$. 
These results are consistent with ED calculations~\cite{Hafezi2007} 
showing that the overlap between the Laughlin state on a grid and the exact ground state increases gradually with increasing $U$ for $\phi=1/3$.  
Finally, we considered a state with a smaller flux density $\phi=1/4$ and 
find that the ratio $R_{\mathcal{I}}$ exhibits a sharp jump at $U/J \approx 9.8$. 
Below and above this critical interaction strength the currents show qualitatively different behaviour, and are similar to the Laughlin state for $U/J > 9.8$. 

In summary, lattice systems exhibit rich physical phenomena in a largely unexplored parameter regime away from  the continuum limit.
\begin{acknowledgments}
MK and DJ acknowledge financial support from the National Research Foundation
and the Ministry of Education, Singapore. DJ acknowledges funding from the 
European Research Council under the European Union's Seventh Framework Programme 
(FP7/2007-2013)/ERC Grant Agreement no. 319286, Q-MAC. 

ML is grateful for funding from the Networked Quantum Information Technologies Hub (NQIT) of the UK National Quantum Technology Programme as well as from the EPSRC grant "Tensor Network Theory for strongly correlated quantum systems" (EP/K038311/1).
\end{acknowledgments}
\appendix
\section{Finite State Automaton} \label{appendixA}

Here we show how we created the MPO that describes our system's Hamiltonian using the finite state automata technique for a sample $L_x\times L_y=3\times 3$ lattice.
A Hamiltonian  $H$ acting on a 1D lattice with $L$ sites is written in the MPO form as
\begin{equation}
H = \!\!\!\sum_{
		j_0,...,j_L\!\!\!}
			v^l_{j_0}M^1_{j_0 j_1}M^2_{j_1 j_2} \ldots M^L_{j_{L-1} j_L}v^r_{j_L}  \;,
\end{equation}
where the $M^i$'s are $d\times d$ matrices whose elements are one-particle operators on site $i$, and ${v}^l$ and ${v}^r$ are vectors.
The finite automata technique consists in deriving the elements $M^i_{ab}$ of the matrix that makes up the MPO at site $i$ from a complex weighed finite automaton. This finite automaton can be thought of as a graph made of nodes and links between the nodes. A node is associated with each index $a$ of the matrix element $M^i_{ab}$. Any two nodes, $a$ and $b$ will have a link between them if the corresponding matrix element $M^i_{ab}$ is non-zero. This non-zero value will be associated with such a link and it represents the weight of the transition between the node $a$ and the node $b$. The set of all non-zero matrix elements can be thought of as creating a path between a starting and finishing node. This automaton is non-deterministic and this means that each node can be connected with more than one other node, and all the different paths will be taken in superposition. All the terms of the Hamiltonian will be obtained by starting at the first node and, after $n$ steps, placing an operator $M^i_{ab}$ on site $n$. The whole Hamiltonian can then be derived by following all the paths that link the starting node to the end node.

\begin{figure}[t]
  \centering
\includegraphics[width=0.4\textwidth]{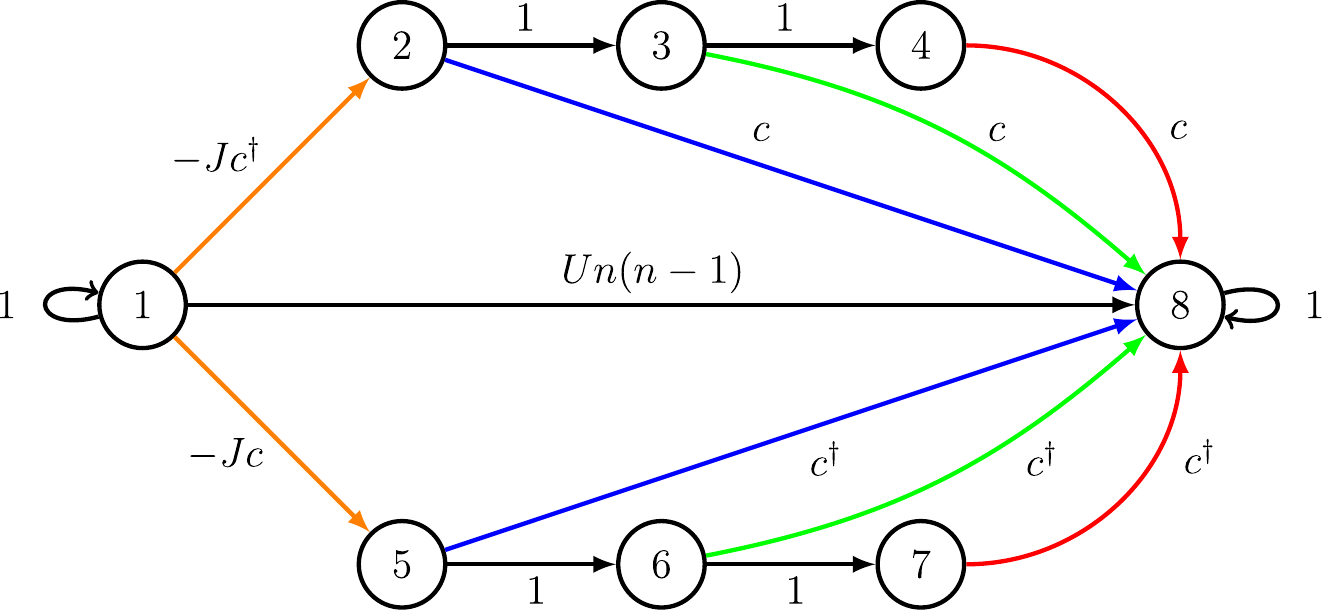}
  \caption{  Finite state automaton used to construct the MPO tensor describing all the terms of the system Hamiltonian for a $3\times 3$ lattice with cylindrical boundary conditions. The coloured arrows represent different terms depending on the position of site $i$ relative to the lattice structure. The blue line represents the NN terms with the same x coordinate in the centre of the lattice. The green line represents the same NN term but around the PBC  and the red lines are the NN terms with the same y-coordinate.  }
  \label{automata_full}
\end{figure}

Looking at the Hamiltonian in Eq. \ref{Ham} we see that there are three types of terms that need to be described by the MPO. The first two are the hopping term and its Hermitian conjugate, and  the third one is the density-density interaction.
The corresponding automaton is shown in Fig. \ref{automata_full}. The three paths branching off from the site labelled with $0$ correspond to the three terms of the Hamiltonian.
Only in the case of a translationally symmetric system would the MPO be made up of copies of the same tensor $M^i$.  In our case, there are two factors that break the translational symmetry. First, the complex hopping in the Landau gauge depends on the $x$ coordinate. This leads to a $t$-site periodicity in $x$ where $\phi=r/t$. Moreover, the tensors of the MPO also need to vary along $y$ because they need to describe different terms at the bottom of the lattice, in the centre and at the top, due to the PBC and the 2D to 1D mapping. 
The red link in Fig. \ref{automata_full} represents the NN terms with the same $y$, the blue one the NN terms with the same $x$ and the green one the PBC term.
The different terms corresponding to the bottom, middle and top position of the sites are reflected by the coefficients added when the $c$ and $c^\dagger$ terms that link to the last node are inserted (for the sake of visual clarity these coefficients have been omitted from Fig. \ref{automata_full}).
We explicitly write the MPO tensor corresponding to a top position in the lattice with the appropriate coefficients as 
\begin{equation}\label{eq:MPO_full}
M^i=
\left(
\begin{array}{c|ccc|ccc|c}
     	1&	-Jc^\dagger	&0	&0	&-Jc	&0	&0	&	 Un(n-1)  \\ \hline
     	  &	&	1	&	&	&	&	&		e^{i2\pi x\phi}c\\
     	  &	&	&	1	&	&	&	&		e^{-i2\pi x\phi}c\\
     	  &	&	&	&	&	&	&		c \\ \hline
     	  &	&	&	&	& 1	 &	&		e^{-i2\pi x\phi}c^\dagger\\
     	  &	&	&	&	&		&1 &	e^{i2\pi x\phi}c^\dagger\\
     	  &	&	&	&	&	&	&		c^\dagger \\ \hline
     	  &	&	&	&	&	&	&		1
	\end{array}
	\right),
	\end{equation}
where for the sake of visual order the empty terms in the matrix represent zeros.
The dimension of the MPO is $2L_y+2$.

\section{Non-interacting ground state} \label{appendixB}
We here describe how to obtain the analytical solution for the ground state of the system on the finite cylinder in the non-interacting limit $U=0$.
We write the non-interacting Hamiltonian as
\begin{multline}\label{Ham_noinnter}
H = -J\bigg[\sum_{x=1}^{L_x-1}\sum_{y=1}^{L_y} c^{\dagger}_{x,y}c_{x+1,y}+\sum_{x=1}^{L_x} \bigg( 
 e^{i2\pi x\phi}c^{\dagger}_{x,L_y}c_{x,1} \\
+ \sum_{y=1}^{L_y-1} e^{i2\pi x\phi} c^{\dagger}_{x,y}c_{x,y+1} \bigg)\bigg]+ \text{H.c.},
\end{multline}
Because of the PBC in the $y$-direction we Fourier transform the bosonic operators as:
\begin{equation}
c_{m,n}=\frac{1}{L_y}\sum_k e^{ikn}c_{m,k},
\end{equation}
where $k=2\pi n_y /L_y$ with $n_y=0,..., L_y-1$ are the discrete values of the momentum in $y$.
In the mixed momentum and real space basis the Hamiltonian becomes
\begin{equation}
H = -J\sum_{x}  \sum_{k} c^{\dagger}_{x,k}c_{x+1,k}+e^{-i2\pi x\phi}c^{\dagger}_{x,k}c_{x,k} + \text{H.c}.
\end{equation}
Let us now consider a one-particle state with momentum $k$
\begin{equation}\ket{\psi(k)}=\sum_m \psi_{m}(k)c^{\dagger}_{mk}\ket{0}.
\end{equation}
By applying the Hamiltonian to the state the Schr\"odinger equation becomes
\begin{multline}\label{Shro}
-J\big(\psi_{m+1}(k)+\psi_{m-1}(k)\big)-2J \cos(k+2\pi\phi m)\psi_{m}(k)=\\
E \psi_{m}(k).
\end{multline}
By using the transfer matrix approach presented in~\cite{Hatsugai1993} we write Eq. \ref{Shro} in matrix form as a recursive relation for the coefficients $\psi_{m}(k)$ as
\begin{equation}\label{recursive}
\begin{pmatrix}
\psi_{m+1}(k, \epsilon)\\ 
\psi_{m}(k, \epsilon)
\end{pmatrix}
=M_m(k, \epsilon)
\begin{pmatrix}
\psi_{m}(k, \epsilon)\\ 
\psi_{m-1}(k, \epsilon)
\end{pmatrix},
\end{equation}
where $\epsilon=E/J$ and 
\begin{equation}
M_m(k, \epsilon)=\begin{pmatrix}
-\epsilon-2\cos(k+2\pi\phi m) & -1\\ 
 1& 0
\end{pmatrix}.
\end{equation}

The parameters of our system are: $L_x=9$, $L_y=6$ and $\phi=1/3$.
Using the recursive relation in Eq.  \ref{recursive} we get
\begin{equation}
\begin{pmatrix}
\psi_{10}(k, \epsilon)\\ 
\psi_{9}(k, \epsilon)
\end{pmatrix}
=\tilde{M}(k, \epsilon)
\begin{pmatrix}
\psi_{1}(k, \epsilon)\\ 
\psi_{0}(k, \epsilon)
\end{pmatrix},
\end{equation}
where 
\begin{multline}
\tilde{M}(k, \epsilon)=M_9(k, \epsilon)M_8(k, \epsilon)\ldots M_1(k, \epsilon)=\\
\begin{pmatrix}
\tilde{M}_{11}&\tilde{M}_{12} \\ 
\tilde{M}_{21} & \tilde{M}_{22}
\end{pmatrix}.
\end{multline}
For a fixed momentum value $k$, and fixing the boundary conditions as $\psi_{0}(k)=\psi_{10}(k)=0$ we obtain the eigenvalues of the problem by solving equation $\tilde{M}_{11}=0$. This is a polynomial of degree $9$ in $\epsilon$ whose roots give the energies of the eigenstates with momentum $k$. The ground state of the system is obtained by solving this equation for each value of $k$ and choosing the smallest root.
Having fixed both $\bar{k}$ and chosen the ground state energy $\bar{\epsilon}$, we get the coefficients $\psi_{m}(\bar{k})$ of the ground state by using the recursive relation of Eq. \ref{recursive}.
We find that for the ground state $\bar{\epsilon}\approx-2.6875$ and $\bar{k}=2\pi/3$.
The ground state $\ket{\bar{\psi}}$ is written in real space as
\begin{equation}
\ket{\bar{\psi}}=\sum_{mn} \bar{\psi}_{m} e^{i2\pi n/3}c^{\dagger}_{mn}\ket{0}
\end{equation}
where the $\bar{\psi}_m$ are such that the ground state function is normalized.
We  use this ground state function to calculate all the relevant observables. We find that
\begin{equation}
\langle n_{mn}\rangle=| \bar{\psi}_{m}|^2,
\end{equation}
\begin{equation}
\mathcal{I}^{y}(m)=2| \bar{\psi}_{m}|^2\sin(\frac{2\pi m}{3}+\frac{2\pi}{3}),
\end{equation}
\begin{equation}
|C^y(m,n)|=1.
\end{equation}

Our system is made of $9$ non-interacting bosons and they will all occupy the one-particle ground state. The many-particle ground state $\ket{\bar{\Psi}}$ is therefore
\begin{equation}
\ket{\bar{\Psi}}=\frac{1}{\sqrt{9}} \big(\sum_{mn} \bar{\psi}_{m} e^{i2\pi/3n}c^{\dagger}_{mn}\big)^9\ket{0}.
\end{equation}
The expectation values for the density and current scale linearly with the number of particles whereas the correlation function does not because it is normalized.
The results of the DMRG calculations are in agreement  with the values obtained from this analysis.

\end{document}